\documentclass[mathleft]{an}
\usepackage{graphicx}
\usepackage{times}
\begin{document}

\Pagespan{10}{}
\Yearpublication{2008}%
\Yearsubmission{2007}%
\Month{7}%
\Volume{329}%
\Issue{1}%
\DOI{10.1002/asna.200710827}
\title{On the distance to the Ophiuchus 
star-forming region}

\author{E.E. Mamajek\inst{}\thanks{Corresponding author:
emamajek@cfa.harvard.edu}}
\titlerunning{On the distance to the Ophiuchus 
star-forming region}
\authorrunning{E.E. Mamajek}
\institute{Harvard-Smithsonian Center for Astrophysics, 
60 Garden St., MS-42, Cambridge, MA 02138, USA}

\received{2007 Jun 6}
\accepted{2007 Jul 30}
\publonline{2007 Dec 28}


\abstract{The Ophiuchus molecular cloud complex has produced in Lynds
1688 the richest known embedded cluster within $\sim$300 pc of the
Sun. Unfortunately, distance estimates to the Oph complex vary by
nearly $\sim$40\% ($\sim$120--165 pc). Here I calculate a new
independent distance estimate of 135\,$\pm$\,8 pc to this benchmark
star-forming region based on Hipparcos trigonometric parallaxes to
stars illuminating reflection nebulosity in close proximity to Lynds
1688. Combining this value with recent distance estimates from
reddening studies suggests a consensus distance of 139\,$\pm$\,6 pc
(4\% error), situating it within $\sim$11\,pc of the centroid of the
$\sim$5 Myr old Upper Sco OB subgroup of Sco OB2 (145 pc).  The
velocity vectors for Oph and Upper Sco are statistically
indistinguishable within $\sim$1\,km\,s$^{-1}$ in each vector
component. Both Oph and Upper Sco have negligible motion
($<$1\,km\,s$^{-1}$) in the Galactic vertical direction with respect
to the Local Standard of Rest, which is inconsistent with the young
stellar groups having formed via the high velocity cloud impact
scenario.}

\maketitle

\section{Introduction}

The Ophiuchus cloud complex, and more specifically the Lynds 1688 dark
cloud, contains the richest embedded cluster within 300 pc of the Sun
(Allen et al. 2002; Lada \& Lada 2003; Porras et al. 2003). Ophiuchus
continues to be critical to our astrophysical understanding of many
aspects of star formation and early stellar evolution (see review by
Wilking, Gagne, \& Allen, in press), including prestellar cloud clumps
(e.g. Motte et al. 1998), the Class O protostellar stage (e.g. Andre
et al. 1993), the dynamics of Class I protostars (e.g. Covey et
al. 2006), substellar objects and substellar binaries (e.g. Luhman et
al. 2007), the initial mass function (e.g. Luhman \& Rieke 1999), and
X-ray emission in T Tauri stars (e.g. Montmerle et al. 1983), among
other topics.

Surprisingly the distance to the complex is not well constrained, with
modern estimates from reddening studies ranging between 120--150 pc
(Knude \& Hog 1998), 125\,$\pm$\,25 pc (de Geus et al. 1989), and
165\,$\pm$\,20 pc (Chini 1981).  A distance estimate to the Oph region
is notably lacking from the studies of Hipparcos distances to nearby
star-forming regions by Wichmann et al. (1998) and Bertout
(1999). While arguments have been made that the Oph cloud is
co-distant with the Upper Sco subgroup of Sco OB2 (145\,$\pm$\,2 pc;
de Zeeuw et al. 1999), it is possible that the group could be in front
of or behind this OB subgroup.  Here I estimate a distance to the
Ophiuchus star-forming region via Hipparcos parallaxes to stars
illuminating reflection nebulosity in the immediate vicinity of the
Oph clouds. Using the improved distance and a new estimate of the mean
proper motion for the Oph stellar group, I show that the motions of
Oph and Upper Sco are currently indistinguishable.

\section{Analysis}

\subsection{Distance}

I queried the Merged Catalogue of Reflection Nebulae (Magakian 2003)
for nebulae within 5$^{\circ}$ in the region of the LDN 1688 cloud,
with the radius chosen to generously sample much of the region where
molecular gas is traced in $^{13}$CO maps (Nozawa et al. 1991). I
identified six reflection nebulosities with at least one bright star
illuminating the structure (vdB 104, 105, 106, 107, 108, and DG
137). The clustered nature of these nebulae are obvious in Fig. 3 of
Magakian (2003), and the rarity of such nebulae at high galactic
latitude argue strongly for their association with the dense gas in
the Oph complex (only $\sim$4\% of the nebulae in the Magakian catalog
are at Galactic latitude $b> 15^{\circ}$).  The Hipparcos
parallaxes (Perryman et al. 1997) for the stars flagged as
illuminating the nebulae (and their companions) are listed in Table
\ref{table:parallax}.

I also queried the Hipparcos catalog with the list of 312 candidate
members of the Oph cloud provided in the review of Wilking et al. (in
press). The only Oph member with a Hipparcos parallax from the Wilking
et al. list is the well-known object HIP 80462 (SR 1), which also
illuminates a reflection nebula. One other object from the Wilking et
al. list had a Hipparcos parallax (HIP 80685 = HD 148352), but its
proper motion and parallax are very large, so I do not consider the
object further in the distance calculation.

\begin{table}
\caption{Hipparcos parallaxes of stars associated with reflection nebulae in Oph
region.}
\label{table:parallax}
\begin{tabular}{llllc}\hline\noalign{\smallskip}
HD     & HIP   & Alias         & Nebula     & $\pi$ (mas)\\
\noalign{\smallskip}\hline\noalign{\smallskip}
147165 & 80112 & $\sigma$ Sco  & vdB 104    & 4.44\,$\pm$\,0.81 \\
147702 & 80377 & SAO 184364    & DG 137     & 4.73\,$\pm$\,1.34 \\
147888 & 80461 & $\rho$ Oph DE & vdB 106    & 7.33\,$\pm$\,1.37 \\
147889 & 80462 & SR 1          & vdB 105    & 7.36\,$\pm$\,1.19 \\
147933 & 80473 & $\rho$ Oph AB & vdB 106    & 8.27\,$\pm$\,1.18 \\
147932 & 80474 & $\rho$ Oph C  & vdB 106    & 7.76\,$\pm$\,0.96 \\
148478 & 80763 & $\alpha$ Sco  & vdB 107    & 5.40\,$\pm$\,1.68 \\
148605 & 80815 & 22 Sco        & vdB 108    & 8.30\,$\pm$\,0.84 \\[0.5mm]
\hline
\end{tabular}
\end{table}

The eight trigonometric parallax values listed in Table
\ref{table:parallax} are consistent with a weighted mean estimate of
$\pi$ = 6.75\, $\pm$\, 0.38 mas, but with a high reduced $\chi^2$
value ($\chi^2$/$\nu$ = 17.7/7 = 2.5). Omitting the most deviant
outlier ($\sigma$ Sco) reduces the $\chi^2$ significantly
($\chi^2$/$\nu$ = 8.0/6 = 1.3) to a value consistent with the range of
expected $\chi^2$ for a good fit (Stuart \& Ord 2005,
Chapt. 16.3). Clipping the next most deviant parallax has negligible effect
($\sim$0.7$\sigma$ level) on the final value. The weighted mean
parallax for the remaining seven stars is $\pi$ = 7.41 \,$\pm$\,
0.43 mas, consistent with a distance of 135.0$^{+8.4}_{-7.4}$ pc (6\%
error).

In Table \ref{table:distances}, I compare the new distance estimate to
previously published values. I combine the new estimate with the
previously published independent values to derive a weighted mean
distance. For Knude \& Hog (1998), I assume that their range of
plausible values (120--150 pc) are consistent with a normal
distribution at 135\,$\pm$\,15 pc. The published distances are
self-consistent ($\chi^2$/$\nu$ = 3.7/3 = 1.2), and lead to a weighted
mean distance of 139\,$\pm$\,6 pc (4\% error). I adopt this as the
best available distance to the Oph star-forming region.

\begin{table}
\caption{Distance estimates to Oph.}
\label{table:distances}
\begin{tabular}{ll}\hline\noalign{\smallskip}
Reference & Dist. (pc)\\
\noalign{\smallskip}\hline\noalign{\smallskip}
Chini 1981                     & 165\,$\pm$\,20 pc\\
de Geus, de Zeeuw \& Lub 1989 & 125\,$\pm$\,25 pc\\
Knude \& Hog 1998              & 120-150 pc\\
This study                       & 135$^{+8}_{-7}$ pc\\
\noalign{\smallskip}\hline\noalign{\smallskip}
Mean                             & 139\,$\pm$\,6 pc\\[0.5mm]
\hline
\end{tabular}
\end{table}

\subsection{Proper motion}

To calculate the velocity vector of the Oph group, one also needs an
estimate of the group's mean proper motion.  Candidate members of the
Oph cloud are generally optically faint and not well represented in
the {Hipparcos} catalog (Perryman et al. 1997), and those that
appear in the Pre-Main Sequence Stars Proper Motion Catalog (Ducourant
et al. 2005) typically have large uncertainties ($\sim$10
mas\,yr$^{-1}$). The candidate Oph members are, however, sufficiently
represented in astrometric catalogs of fainter stars (Tycho-2, UCAC2,
SPM2.0) that one can estimate the mean proper motion of the stellar
group in LDN 1688. I queried the 312 candidate Oph cloud members from
Wilking et al. with entries in the Tycho-2 catalog (H{\o}g et
al. 2000, 4 matches), UCAC2 catalog (Zacharias et al. 2004, 12
matches), and SPM-2.0 catalog (Platais et al.  2007, 35 matches). The
proper motions and astrometric aliases for these Oph candidate members
are listed in Table \ref{table:proper_motions}.

Through cross-referencing entries in the UCAC2 and SPM2.0 catalogs
with those in the Tycho-2 catalog (all matches within 2$^{\circ}$ of
the center of the Oph stellar group), I find that UCAC2 and SPM2.0
proper motions are consistent with being on the Tycho-2 system within
$<$1 mas\,yr$^{-1}$ (where Tycho-2 is tied to the inertial ICRS at the
$\sim$0.25 mas\,yr$^{-1}$ level; H{\o}g et al. 2000). Among the 53
stars with a counterpart in one of the astrometric catalogs, the
proper motion with the smallest uncertainty was selected.  Calculation
of the mean proper motion (and its uncertainty) for the group were
made using three estimators which are fairly insensitive to outlying
points, including the true median (Gott et al. 2001) and the clipped
mean using Chauvenet's criterion (Bevington \& Robinson 1992).  The
proper motions were also run through a custom-made clipping routine
which iteratively clips the most discrepant outliers until a mean
value is found that gives a $\chi^2$ sufficiently low to be considered
a good fit considering the degrees of freedom (Stuart \& Ord
2005). For the clipping routine, an internal velocity dispersion of
1.5 mas\,yr$^{-1}$ was assumed, appropriate for 1\,km\,s$^{-1}$
(typical for young clusters) at $d$ = 139 pc.  I justify the clipping
of the data outliers on the grounds that our parent sample may contain
interlopers unrelated to the Oph cloud (e.g. HD 148352) and probable
cloud members whose motions are likely perturbed due to binarity
(e.g. SR 1). The three estimators converge on a mean proper motion of
($\mu_{\alpha {\rm cos}\,\delta}, \mu_{\delta}$ = --10, --27
mas\,yr$^{-1}$). The statistical error in each component of this
estimate is $\sim$1.5 mas\,yr$^{-1}$, and the estimated systematic
errors are $\sim$1 mas\,yr$^{-1}$ (for the UCAC2 and SPM catalogs, on
which this analysis is heavily dependent), so I conservatively assign
a total uncertainty in the mean proper motion of 2 mas\,yr$^{-1}$. The
mean proper motion of the Oph cloud members is remarkably close to the
median value for 120 members of the adjacent Upper Sco group:
($\mu_{\alpha {\rm cos}\,\delta}, \mu_{\delta}$ = --11, --24
mas\,yr$^{-1}$, de Zeeuw et al. 1999).

An investigation of the kinematics of individual candidate group
members is beyond the focus of this study, but one should note that
some Oph candidates have large proper motions clearly inconsistent
with group membership (especially HD 148352, [GY92] 165, and [GY92]
280).  The {Hipparcos} astrometry and photometry of the high
proper motion star HD 148352 are consistent with it being an
unreddened foreground F dwarf, unrelated to the Oph cloud. The 2MASS
and SPM photometry of the high proper motion star [GY92] 165 are
consistent with it being an unreddened foreground late-K dwarf, rather
than a member of the Oph cloud. The high proper motion star [GY92] 280
has 2MASS photometry consistent with an unreddened early-M
dwarf. Indeed Wilking et al. (2005) classify the star as an M2
``dwarf?'' and Strom et al. (1995) classify it as ``foreground'', so
we confirm its interloper status based on its proper motion.  Neither
[GY92] 165 \& 280 have been detected in deep X-ray images, again
supporting their status as old interlopers.  All three should be
rejected from Oph membership lists.  Two stars with accurate proper
motions that were consistently clipped (SR 2 \& 9) are likely to be
bona fide Oph members whose photocentric motion is perturbed by their
binarity.

\subsection{Space velocity and position}

I calculate the Galactic space motion vector for the members of the
Oph cloud using the best estimate of the distance (139\,$\pm$\,6 pc),
proper motion ($\mu_{\alpha {\rm cos}\,\delta}, \mu_{\delta}$ =
--10\,$\pm$\,2, --27\,$\pm$\,2 mas\,yr$^{-1}$), the median position of
the Wilking et al. cloud members ($\alpha, \delta$ = 246.78$^{\circ}$,
--24.48$^{\circ}$), and the median RV for Oph pre-MS stars
(--6.3\,$\pm$\,0.3 km\,s$^{-1}$ Prato 2007; Guenther et al. 2007;
Kurosawa, Harris \& Littlefair 2007; James et al. 2006).  The
Galactic space motion vector (velocities with respect to the Sun, no
correction for Galactic rotation) is ($U, V, W$ = --6.2, --17.1, --8.3
km\,s$^{-1}$) with errors in the velocity components of (1.0, 1.3, 1.3
km\,s$^{-1}$). Given that the Sun's peculiar motion in the Z direction
with respect to the Local Standard of Rest (LSR) is +7.2\,$\pm$\,0.4
km\,s$^{-1}$ (Dehnen \& Binney 1998), I find that Oph is statistically
consistent with having negligible vertical motion with respect to the
LSR ($\Delta W$ = --1.2\,$\pm$\,1.3 km\,s$^{-1}$).

From Monte Carlo simulations that sample the uncertainties in the best
estimates for the distances and centroid positions for Oph and the
Upper Sco OB subgroup of Sco OB2 (145 pc; de Zeeuw et al. 1999), I
find that the group centroids are only 11\,$\pm$\,3 pc in separation,
with Oph slightly in the foreground.  The centroid position of the Oph
cloud population, in Galactic coordinates ($X$ towards Galactic
center, $Y$ in direction of Galactic rotation, $Z$ towards the North
Galactic Pole) is ($X, Y, Z$ = 132, --16, 40 pc). The median position
of the Upper Sco population (de Zeeuw et al. 1999) is ($X, Y, Z$ =
135, --21, +49 pc). The properties of the Oph stellar group are
summarized in Table \ref{table:properties}. The relative positions
agree with de Geus's (1992) picture where the $\rho$ Oph cloud is
situated in the foreground of the slightly more distant Upper Sco OB
subgroup.

The massive stars in Upper Sco have often been cited as the agent
responsible for triggering the star-formation in the Oph cloud
(e.g. de Geus 1992; Preibisch et al. 2002), so it is of interest to
models of triggered star-formation what the bulk velocity of the Oph
cloud membership is with respect to the Upper Sco subgroup. Although
the Upper Sco subgroup is well-studied, surprisingly there is an
uncomfortably large range of Galactic space motion velocity estimates
in the recent literature (de Zeeuw et al. 1999; de Bruijne 1999;
Madsen et al. 2002). Thus far the velocity estimates of the centroid
motion for Upper Sco are critically dependent on the location of the
convergent point and the tangential velocity for estimating the
Galactic velocity components.  However, for an {\it expanding}
subgroup, the convergent point analysis will give the velocity of a
group member participating in the linear expansion at the position of
the Sun (e.g. Brown et al. 1997; Mamajek 2005), which is not the
quantity that interests us here.  That the published velocity vectors
for the Sco-Cen subgroups inferred from convergent point analysis
alone (i.e. ignoring radial velocity data; e.g. Madsen et al. 2002)
are probably in error is demonstrated by the fact that the vectors
predict radial velocities that are systematically off of the mean
observed RV by several km\,s$^{-1}$.  Including an estimate of the
group radial velocity in the analysis gives a more accurate estimate
of a group's bulk motion, independent of the effects of any
expansion. To estimate the bulk motion of Upper Sco I calculate the
velocity vector using the median position, distance, proper motion,
and radial velocity for the 120 Upper Sco members from de Zeeuw et
al. (1999). This leads to a space velocity of ($U, V, W$ = --5.2,
--16.6, --7.3 km\,s$^{-1}$), with component errors of only $\sim$0.3
km\,s$^{-1}$. As with Oph, Upper Sco demonstrates negligible motion in
the vertical direction with respect to the LSR ($\Delta W$ =
--0.1\,$\pm$\,0.5 km\,s$^{-1}$).  Their relative motion (in the sense
Oph minus Upper Sco) is then ($\Delta U$, $\Delta V$, $\Delta W$) =
(--1.0, --0.6, --1.1) $\pm$ (1.0, 1.3, 1.3) km\,s$^{-1}$. Hence, Oph
is moving at a negligible 1.6\,$\pm$\,2.1 km\,s$^{-1}$ with respect to
the Upper Sco population. {\it These results show that the velocities
of Oph and Upper Sco are statistically indistinguishable at the
$\sim$km\,s$^{-1}$ level, and that both groups have negligible
vertical motion with respect to the LSR}.

\begin{table}
\caption{Properties of the Lynds 1688 group.}
\label{table:properties}
\begin{tabular}{lc}
\hline\noalign{\smallskip}
Property & Value\\
\noalign{\smallskip}\hline\noalign{\smallskip}
Distance                    & 139\,$\pm$\,6 pc\\
Distance Modulus            & 5.72\,$\pm$\,0.09 mag\\
$\overline{\mu_{\alpha {\rm cos}\delta}}$ & --10\,$\pm$\,2 mas\,yr$^{-1}$\\
$\overline{\mu_{\delta}}$   & --27\,$\pm$\,2 mas\,yr$^{-1}$\\
Radial Velocity             & --6.3\,$\pm$\,0.3 km\,s$^{-1}$\\
Position (ICRS)             & 246\fdg78, --24\fdg48\\
Position (Galactic)         & 353\fdg11, +16\fdg74\\
Position ($X,Y,Z$)          & 132, --16, 40 pc\\
Velocity ($U,V,W$)          & --6.2, --17.1, --8.3 km\,s$^{-1}$\\[0.5mm]
\hline
\end{tabular}
\end{table}

\begin{table}
\caption{Proper motions of candidate Oph members.}
\label{table:proper_motions}
\resizebox{\linewidth}{!}{
\begin{tabular}{llrr}
\hline\noalign{\smallskip}
Name & Astrometric & $\mu_{\alpha {\rm cos}\,\delta}$~~ & $\mu_{\delta}$~~~~~~~\\[0.6mm] 
     & Alias       & (mas\,yr$^{-1}$)   & (mas\,yr$^{-1}$)\\
\noalign{\smallskip}\hline\noalign{\smallskip}
BKLT    J162643-241112*  &  U 22028257   &  --8.2\,$\pm$\,7.8   &  --23.2\,$\pm$\,5.0\\
DoAr    21*              &  S 57602749    &  --20.5\,$\pm$\,3.6  &  --21.0\,$\pm$\,3.6\\
DoAr    24*              &  S 57602752    &  --7.6\,$\pm$\,3.6   &  --26.3\,$\pm$\,3.7\\
DoAr    24E*             &  S 57602750    &  --6.8\,$\pm$\,3.7   &  --29.4\,$\pm$\,3.8\\
DoAr    25               &  U 21797656   &  --10.4\,$\pm$\,5.0  &  --23.2\,$\pm$\,5.0\\
DoAr    25*              &  S 57602745    &  --12.9\,$\pm$\,3.6  &  --27.4\,$\pm$\,3.6\\
DoAR    32*              &  S 57603012    &  --5.8\,$\pm$\,4.2   &  --22.9\,$\pm$\,4.2\\
DoAR    33*              &  S 57603011    &  --9.1\,$\pm$\,3.8   &  --27.2\,$\pm$\,3.8\\
HD      148352           &  S 57603135    &  --52.6\,$\pm$\,3.3  &  --64.7\,$\pm$\,3.4\\
HD      148352           &  T 6799 930 1  &  --49.9\,$\pm$\,1.5  &  --63.1\,$\pm$\,1.5\\
HD      148352*          &  H 80685       &  --52.3\,$\pm$\,0.9  &  --64.2\,$\pm$\,0.7\\
ROX     2*               &  S 57602512    &  --10.5\,$\pm$\,5.3  &  --37.9\,$\pm$\,5.3\\
ROX     3*               &  S 57602744    &  --13.0\,$\pm$\,5.9  &  --28.0\,$\pm$\,6.1\\
ROX     4*               &  S 57602746    &  --5.6\,$\pm$\,4.7   &  --18.7\,$\pm$\,4.7\\
ROX     7*               &  S 57602748    &  --33.0\,$\pm$\,6.9  &  --13.0\,$\pm$\,6.9\\
ROX     16*              &  S 57602906    &  --17.1\,$\pm$\,6.0  &  --26.6\,$\pm$\,6.0\\
ROX     31               &  S 57603002    &  --9.9\,$\pm$\,7.0   &  --27.5\,$\pm$\,7.0\\
ROX     31*              &  U 21797684   &  --12.0\,$\pm$\,5.3  &  --17.1\,$\pm$\,5.3\\
RX      J1624.9-2459*    &  U 21797634   &  --10.2\,$\pm$\,5.3  &  --1.9\,$\pm$\,5.5\\
SR      1                &  H 80462       &  --2.3\,$\pm$\,1.4   &  --25.5\,$\pm$\,1.0\\
SR      1                &  S 57602509    &  --3.1\,$\pm$\,3.5   &  --27.0\,$\pm$\,3.5\\
SR      1*               &  T 6798 539 1  &  --1.4\,$\pm$\,0.9   &  --25.0\,$\pm$\,1.0\\
SR      2                &  S 57602510    &  --18.5\,$\pm$\,3.1  &  --25.2\,$\pm$\,3.3\\
SR      2                &  T 6798 544 1  &  --20.2\,$\pm$\,2.4  &  --27.4\,$\pm$\,2.2\\
SR      2*               &  U 22028253   &  --20.5\,$\pm$\,1.9  &  --26.2\,$\pm$\,1.4\\
SR      3                &  U 21797645   &  --9.7\,$\pm$\,9.8   &  --33.9\,$\pm$\,5.3\\
SR      3*               &  S 57602747    &  --14.3\,$\pm$\,3.8  &  --29.0\,$\pm$\,3.6\\
SR      4                &  S 57602751    &  --14.5\,$\pm$\,6.3  &  --27.8\,$\pm$\,6.3\\
SR      4*               &  U 22028255   &  --14.1\,$\pm$\,5.1  &  --17.7\,$\pm$\,5.0\\
SR      8*               &  S 57602506    &  --7.5\,$\pm$\,3.6   &  --26.7\,$\pm$\,3.6\\
SR      9                &  S 57603009    &  --13.3\,$\pm$\,3.3  &  --30.6\,$\pm$\,3.7\\
SR      9                &  T 6794 513 1  &  --10.5\,$\pm$\,3.2  &  --32.3\,$\pm$\,3.0\\
SR      9*               &  U 22028258   &  --15.5\,$\pm$\,1.5  &  --33.5\,$\pm$\,1.5\\
SR      10*              &  S 57603007    &  --9.0\,$\pm$\,3.6   &  --26.6\,$\pm$\,3.7\\
SR      12*              &  S 57603001    &  --9.8\,$\pm$\,7.5   &  --30.8\,$\pm$\,6.0\\
SR      13*              &  S 57603136    & --9.6\,$\pm$\,3.6   &  --28.5\,$\pm$\,3.6\\
SR      20*              &  S 57603138    &  --8.2\,$\pm$\,3.7   &  --27.8\,$\pm$\,3.7\\
SR      21*              &  S 57602905    &  --7.3\,$\pm$\,3.6   &  --33.0\,$\pm$\,3.7\\
SR      22               &  S 57602508    & --10.4\,$\pm$\,6.1  &  --24.0\,$\pm$\,6.1\\
SR      22*              &  U 22028254   &  --5.2\,$\pm$\,5.5   &  --25.4\,$\pm$\,5.0\\
SR      24*              &  S 57602902    &  --4.9\,$\pm$\,6.2   &  --23.4\,$\pm$\,6.2\\
WSB     28*              &  S 57602753    &  --18.5\,$\pm$\,7.2  &  --22.9\,$\pm$\,7.2\\
WSB     40*              &  S 57602907    &  --21.5\,$\pm$\,6.8  &  --19.5\,$\pm$\,6.8\\
WSB     45*              &  S 57602897    &  --0.5\,$\pm$\,7.0   &  --36.8\,$\pm$\,7.0\\
WSB     46               &  U 21797671   &  --12.0\,$\pm$\,5.3  &  --25.1\,$\pm$\,5.0\\
WSB     46*              &  S 57602896    &  --12.6\,$\pm$\,4.4  &  --30.8\,$\pm$\,4.4\\
WSB     48*              &  S 57602997    &  --11.2\,$\pm$\,6.8  &  --31.4\,$\pm$\,6.8\\
WSB     49*              &  U 21797676   &  --7.7\,$\pm$\,5.2   &  --22.0\,$\pm$\,5.1\\
$[$GY92$]$ 112           &  S 57602904    &  --3.3\,$\pm$\,7.6   &  --42.2\,$\pm$\,7.6\\
$[$GY92$]$ 112*          &  U 21797666   &  1.9\,$\pm$\,5.2    &  --18.3\,$\pm$\,5.0\\
$[$GY92$]$ 165*          &  S 57602901    &  51.8\,$\pm$\,3.6   &  --60.9\,$\pm$\,3.6\\
$[$GY92$]$ 280*          &  S 57603003    &  78.3\,$\pm$\,7.3   &  --56.8\,$\pm$\,7.3\\
$[$GY92$]$ 372*          &  S 57603008    &  --8.5\,$\pm$\,4.4   &  --31.4\,$\pm$\,4.4\\[0.5mm]
\hline
\end{tabular}
}
\\[1.5mm]
Notes: ``*'' flags the proper motion with smallest uncertainty for
each star. ``H'' is HIP, ``U'' is UCAC2, ``S'' is SPM2.0, ``T'' is
Tycho-2.  ``SR'' appears in SIMBAD as ``Em* SR''.
\end{table}

\section{Discussion}

Our independent distance to Oph (135 \,$\pm$\, 8 pc) is comfortably
within the range of previous distance estimates (120--166 pc). The
distance to Oph is very similar to that of other star-forming clouds
in its vicinity, including the neighboring Pipe Nebula $\sim$35\,pc to
the east ($d$ = 130$^{+13}_{-20}$ pc, Lombardi et al. 2006), the Lupus
cloud complex $\sim$35\,pc to the west ($d$ = 140\,$\pm$\,20 pc;
Hughes et al. 1993), and the Corona Australis complex $\sim$80\,pc to
the south ($d$ = 129\,$\pm$\,11 pc; Casey et al. 1998).  These
clouds are also in close proximity to the Upper Sco and Upper Cen-Lup
OB subgroups (both at $d$ $\simeq$ 140\,pc, with ages $\sim$5 and
$\sim$15 Myr, respectively), suggesting that these star-forming clouds
and the Sco-Cen OB association formed from the same large-scale
process.

The similarity in velocities between the $\sim$5 Myr Upper Sco members
and the newly formed Oph members suggests that Upper Sco and Oph
formed from gas with roughly the same bulk motion.  The kinematic data
can be used to discount the high velocity cloud (HVC) impact model for
forming Ophiuchus. Lepine \& Duvert (1994) proposed that the Oph cloud
(and the entire Sco-Cen complex) was formed as the result of a HVC
impact, where the progenitor HVC impacted the Galactic plane at
$\sim$250 km\,s$^{-1}$. In order to explain the distribution of
positions and ages of young stars in the Oph-Sco-Cen region, the Oph
cloud (representing the dense, shocked layer in the collision) was
predicted to be falling towards the Galactic plane from a maximum
height of $Z$ $\simeq$ +100\,pc (where it would have negligible
vertical motion). However, both Oph and Upper Sco appear to have
negligible motion in the $Z$ direction with respect to the LSR at
their current locations. The scenario also predicts that the
velocities of young stars formed in the HVC impact should have
significantly different velocities as the shocked layer is decelerated
and the motions of the newly-formed stars are dominated by the
Galactic potential rather than the motions of the gas. The similarity
of the vertical motions of Oph and Upper Sco (within $\sim$1
km\,s$^{-1}$ of each other and the LSR) appear to also be inconsistent
with this prediction.

The kinematic data and star-formation history of the Oph and Upper Sco
region give us some clue regarding the nature of the two older
($\sim$15 Myr), and more distended Sco-Cen subgroups: Upper Cen-Lup
(UCL) and Lower Cen-Cru (LCC). Although larger and older than Upper
Sco, UCL and LCC have similar velocity dispersions as Upper Sco
($\sim$1 km\,s$^{-1}$, de Bruijne 1999; Mamajek, Meyer \& Liebert
2002).  Despite the km\,s$^{-1}$-level coherence in the motions of its
members, LCC shows some evidence of an age spread with the northern
part of the group having mean age $\sim$17 Myr, while the southern
part (the Southern Cross) has mean age $\sim$12 Myr (Preibisch \&
Mamajek, in press). One can imagine that Upper Sco ($\sim$5 Myr age)
and Oph ($<$2 Myr) may evolve into a LCC-like configuration in
$\sim$10 Myr time, after the newly formed massive stars have cleared
the Oph region of its star-forming molecular gas.  If the Oph young
stellar population is unbound after its molecular gas is dispersed,
then a future observer of the Sco-Oph region (say $\sim$10 Myr in the
future) with kinematic information of $\sim$km\,s$^{-1}$ accuracy
would have difficulty disentangling the members of Oph and Upper Sco
by any kinematic criteria. Following our observations of the Oph and
Upper Sco regions, it seems possible that UCL and LCC are each
comprised of the unbound remnants of multiple embedded clusters with
similar bulk motions (within $\sim$1\, km\,s$^{-1}$) that formed over
a $<$10 Myr span, rather than the remnants of two large embedded
clusters that formed in single bursts.

The Oph cloud is being impacted by the Upper Sco bubble of atomic
hydrogen and molecular gas, presumably the remnants of the proto-Upper
Sco molecular cloud (de Geus 1992).  If one hypothesizes that Oph
represents a long-lived remnant clump of the proto-Upper Sco cloud,
then it has apparently inherited $<$2 km\,s$^{-1}$ of relative
velocity from the expansion of the Upper Sco bubble. The kinematic
data are also inconsistent with the idea that the Upper Sco stars
formed from the {\it contemporary} Oph clouds, and ``migrated'' to
their current positions. The data are consistent with the idea that
the Oph cloud complex and Upper Sco proto-cloud formed from the same
large scale process, which endowed them with similar velocities and
positions in close proximity ($\sim$10 pc), but that conditions for
star-formation in Upper Sco were ripe (and then soon extinguished)
$\sim$5 Myr before that in Oph.

\acknowledgements

EM is supported by a Clay Postdoctoral Fellowship from the Smithsonian
Astrophysical Observatory.

\section{Note Added In Proof}
Floor van Leeuwen (2007, {\it Hipparcos, the New Reduction of the Raw
Data}, Springer) has recently published revised trigonometric
parallaxes from the Hipparcos data for the stars in Table 1.
Repeating the calculation in \S2.1 using the new parallaxes results in
a mean parallax of $\pi$ = 7.62\, $\pm$\, 0.15 mas (131 \,$\pm$\, 3 pc
; distance modulus = 5.59\, $\pm$\, 0.04 mag). Rejecting $\sigma$ Sco
(HIP 80112) again improves the solution from $\chi^2/\nu$ = 32.3/7 to
6.7/6 (= 1.1). The revised distance of 131\, $\pm$\, 3 pc (2\%\,
error) is probably the best available derived from Hipparcos data. The
revised mean velocity vector for the Oph group is ($U, V, W$ = --6.2,
--16.1, --8.0 km\,s$^{-1}$) with errors in the velocity components of
(0.9, 1.1, 1.2 km\,s$^{-1}$).  This makes its relative motion with
respect to Upper Sco slightly smaller (1.3\, $\pm$ 1.9
km\,s$^{-1}$). The slight shift in distance has negligible impact on
both the quantitative and qualitative conclusions of this study.

\end{document}